# A new system for evaluating brand importance: A use case from the fashion industry

Fronzetti Colladon, A., Grippa, F., & Segneri, L.

# A new system for evaluating brand importance: A use case from the fashion industry

Andrea, Fronzetti Colladon*
Department of Engineering, University of Perugia, Perugia, Italy, andrea.fronzetticolladon@unipg.it

Francesca, Grippa
College of Professional Studies, Northeastern University, Boston, MA, USA, f.grippa@northeastern.edu

Ludovica, Segneri
Department of Engineering, University of Perugia, Perugia, Italy, ludovica.segneri@collaboratori.unipg.it

**Abstract**

Today brand managers and marketing specialists can leverage huge amount of data to reveal patterns and trends in consumer perceptions, monitoring positive or negative associations of brands with respect to desired topics. In this study, we apply the Semantic Brand Score (SBS) indicator to assess brand importance in the fashion industry. To this purpose, we measure and visualize text data using the SBS Business Intelligence App (SBS BI), which relies on methods and tools of text mining and social network analysis. We collected and analyzed about 206,000 tweets that mentioned the fashion brands Fendi, Gucci and Prada, during the period from March 5 to March 12, 2021. From the analysis of the three SBS dimensions - prevalence, diversity and connectivity - we found that Gucci dominated the discourse, with high values of SBS. We use this case study as an example to present a new system for evaluating brand importance and image, through the analysis of (big) textual data.

*Corresponding author.

## 1   The Semantic Brand Score and Its Web App

In this paper, we describe an application of the Semantic Brand Score (SBS) to luxury brands. The SBS is a methodology of assessment of brand importance that combines methods and tools of text mining and social network analysis [1]. We show how to use the Semantic Brand Score Business Intelligence App, which produces a wide range of text and brand analytics, and also allows the automatic mining of (big) textual data - thus helping counteract biases introduced by lack of accuracy in traditional data collection methods such as polls, surveys and focus groups.

The SBS can be calculated on any collection of text documents, such as newspaper articles, emails, or tweets; it can be adapted to multiple languages and applied to "brands" in a broad sense – for example to politicians' names. The SBS has already proved very useful for tourism, financial and election forecasting. See for example [2, 3].

Gaining a deeper understanding of semantic importance and image can change the way we make decisions and manage organizations in the era of big data. The SBS BI App relies on methods and tools of text mining and social network analysis. Its analytical power extends beyond commercial brands, including products and services, personal brands, keywords representing values or concepts associated with societal trends. The methodological approach has been used to build predictive models to explore tourism trends, select advertising campaign testimonials, and make economic, financial and political forecasts.

SBS BI is a tool for discourse and brand analysis that looks at the relationships among words. For this reason, the calculation of the SBS is based on the construction of a word co-occurrence network, and goes beyond the mere count of word frequencies. Building this network, it is possible to get an immediate representation of the brand associations in the discourse and of their strength.

The app is also capable of modeling the main discourse topics, again using a network approach [4, 5], to offer insights about the main themes related to a brand. Topics are identified through network clustering [6] and the most relevant words of each cluster are selected based on their weighted degree and on the proportion of internal and external links [7]. The analysis of the main discourse topics identifies clusters of representative words that can be associated with the brands or exist independently from them. This complements the visualization of "brand associations", which depicts only direct links between words.

Another level of analysis, offered by the SBS BI app, is the ability to look at how brands are related to words that regard several language dimensions – such as the "affective" dimension, the "masculine" and "feminine" dimensions, or those dimensions that identify a language focus on the present, future or past [8].

The SBS BI app is also able to compute other sophisticated measures of language complexity and informativeness. For example, the app takes into account whether the words in a tweet are common or rare, with respect to the overall discourse. To this purpose, it uses a Term Frequency-Inverse Document Frequency (TF-IDF) logic [9] and calculates a metric of novelty of the tweet content as:

$$Novelty = \frac{1}{n}\sum_{w \in V} f_w \log \frac{N}{n_w} \qquad (1)$$

where *N* is the total number of tweets in the corpus; *n* is the total number of words that appear in a tweet and *V* is their set; $f_w$ is the frequency of word *w* and $n_w$ is number of tweets where the word w appears. Taking the sum of tf-idf scores, instead of their average, is also a feasible alternative, in order to calculate the informativeness of each tweet. This works well when the documents in the corpus have comparable lengths. The novelty metric is otherwise useful to understand if some novel content is buried under a lot of redundant information.

In Table 1, we briefly recall the SBS components, which together make the composite indicator. The metric is calculated after text pre-processing and considering networks of co-occurring words [1].

**Table 1:** SBS components.

| SBS component | Description |
|---|---|
| **Prevalence** | It represents the frequency with which a brand name appears in a set of text documents: the more frequently users mention a brand across texts, the higher its prevalence. |
| **Diversity** | It measures the heterogeneity of the words co-occurring with a brand, assigning higher "lexical" diversity to brands embedded in a richer discourse. The distinctiveness centrality measure is used to calculate diversity [10]. |
| **Connectivity** | It measures how often a brand serves as an indirect link between all the other pairs of words. It provides evidence of the connective power of brands, i.e. the ability to connect different words. Connectivity is calculated using weighted betweenness centrality [11]. |

## 2 An Application to Fashion Brands

We collected 206,000 tweets that mentioned three fashion brands - Fendi, Gucci and Prada - in the week from March 5 to March 12, 2021. Table 2 provides the descriptive statistics of our text corpus. By using the SBS BI app, we preprocessed the corpus to remove stop-words and word affixes (stemming).

**Table 2.** Corpus descriptive statistics.

| Measure | M | SD |
|---|---|---|
| Number of Words (Tokens) | 18.73 | 10.81 |
| Number of Unique Words (Types) | 17.05 | 9.05 |
| Type/Token Ratio | 0.93 | 0.07 |
| Six-Letter Words | 29% | 13% |

We downloaded tweets of an average length of 19 words, most of which are not repeated within the single tweet. About 29% of words is made of six letters or longer, which we use as a possible indicator of language complexity [8].

### 2.1 Results

Figure 1 shows the SBS time trends for the three brands. Gucci is the one with the highest SBS score (235.47 on average), characterized by a peak on March 10 (282.3). Prada and Fendi score significantly lower than Gucci, with an average SBS equal to 35.67 and 16.69 respectively. Notwithstanding the positive trends, there is a significant difference with Gucci, which is 7 times higher than Prada and 14 times higher than Fendi.

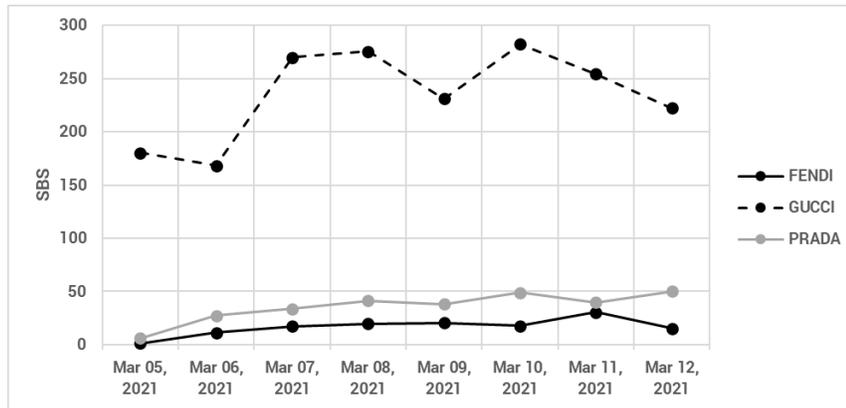

**Figure 1.** SBS trends over time.

From the analysis of the three SBS dimensions (Figure 2), we see that Gucci dominates the discourse, with maximum values of prevalence, diversity and connectivity. This depends in part on the Twitter search query that we used for the case study, which directly included the brand names. In future analyses it may be useful to structure more general queries, for example using terms related to product categories or business sectors, and then study the importance of the target brands.

An interesting insight is that Fendi's and Gucci's prevalence scores are lower than their diversity, implying that, despite the richness of the associations of the two brands, the frequency with which they are mentioned is rather low. The connectivity of Prada and Fendi is also low, suggesting that the two brands are not at the core of the online discussion during the one-week period.

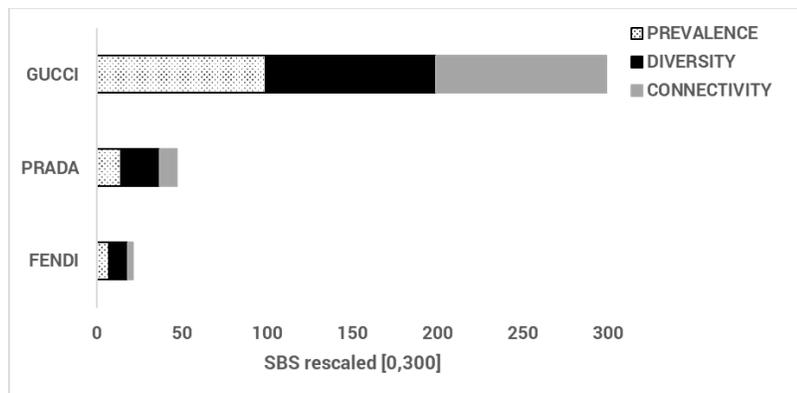

**Figure 2.** SBS score dimensions for Gucci, Prada and Fendi.

If we consider brand associations, we find that their sentiment is much more positive for Gucci (0.27) than for Fendi (0.09) or Prada (0.07), for which it is still positive but closer to neutral. This is aligned with the 2021 study of online popularity of luxury brands conducted by the digital marketing agency LUXE Digital, which found that Gucci remains the number one most popular luxury brand online with 15.2% of the total search interest for luxury goods, well ahead of its competitors for four consecutive years [12].

The study of brand associations and brand topics is important as it allows exploring in detail what drives the importance scores calculated with the SBS. Figure 3 illustrates the direct associations.

During the one-week observation period, Gucci's image is tightly connected with the online discussion about the South Korean singer Kai, elected ambassador of the brand. Kai is member of the band EXO, chosen by Gucci's creative director for a new capsule collection. The collection is focused around an iconic symbol in the collective mind, i.e. a teddy bear.

The collection was launched exclusively at the Gucci store located in Singapore's Ion Orchard shopping mall. In addition, the brand is associated with the name of two artists, Adam Driver and Lady Gaga, starring in the film directed by Ridley Scott entitled "House of Gucci". The film was shot in Italy and tells the story of the murder of Maurizio Gucci that involved his ex-wife, Patrizia Reggiani. The presence of the word "bag" among the top associations suggests that one of the most identifiable products of the Gucci collection is the handbag. Considering both Gucci's score on the SBS components and its associations with other words, we could advance the hypothesis that the engagement of Kai and the choice to support the film "House of Gucci" were functional to the increase of brand importance.

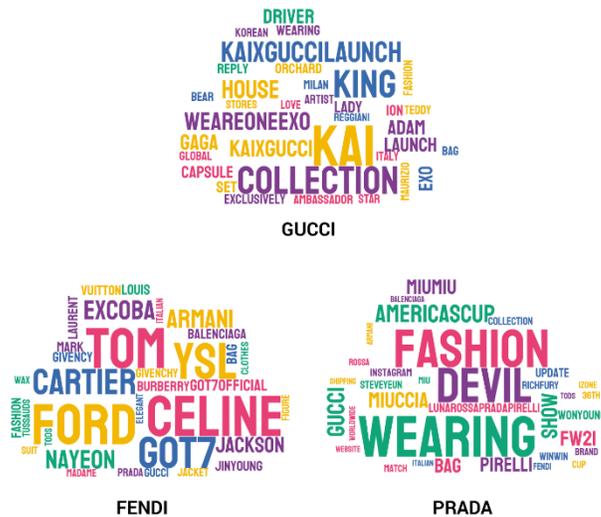

**Figure 3.** Top brand associations.

Among the three brands, Gucci is the only one that does not have the name of its competitors, Fendi and Prada, among the main associations. Given its larger popularity in terms of online search share [12], we might speculate that Fendi and Prada are often referred back to Gucci as term of comparison in the online discourse.

Fendi, on the other hand, is associated with many competitors in the fashion industry such as Tom Ford, Armani, Burberry, YSL, and Celine, as well as with the name of specific Twitter accounts. These associations are less distinctive and less focused than Gucci's associations, and might suggest that Fendi is frequently mentioned in general tweets related to the fashion industry, i.e. together with its competitors. This result is also influenced by its low value of prevalence during the observation period. Interestingly, Fendi has the word "elegant" among its top associations, which attributes a positive quality to the image of the brand.

With regard to Prada, we see that the discourse about this brand is partially mixed with that of the famous movie "The Devil Wears Prada", which generated a positive buzz around the brand. We also find words related to the America's Cup, such as "rossa", "Pirelli" and "lunarossapradapirelli". Indeed, Prada was the presenting sponsor of the event and together with Pirelli supported the Italian boat, Luna Rossa. Moreover, we find Miu Miu among Prada's associations, and other fashion brands such as Balenciaga, Tods, Armani, Fendi and Gucci. The image of Prada is not particularly linked to a specific product, even though the association with the generic word "bag" implies that that term is among the most relevant, at least on Twitter.

In Table 3, we present five topics that emerged from the analysis and in Figure 4 we show how much each brand is associated to each topic.

Table 3. Main discourse topics.

| Topic | Relevance | Top keywords |
| --- | --- | --- |
| Collaboration Gucci-Kai | 49% | Kai, kaixguccilaunch, collection, weareoneexo, kaixgucci, launch, Exo, Capsule, korean, bear, celebrity, teddy, ambassador, artist |
| Fashion brands | 10% | Fendi, Tom Ford, Celine, YSL, Armani, Cartier, Givenchy, Balenciaga, Louis Vuitton, Ralph Lauren, Burberry |
| "House of Gucci" Film | 24% | Lady Gaga, Adam Driver, Ridley Scott, House, set, Patrizia Reggiani, Maurizio Gucci, filming, Italy, Milan, star, playing, spoiler |
| Events | 8% | fw21, show, miumiufw21, fashion, miusassy, lunarossapradapirelli, pirelli, update, match, cup, rossa, race, 36th, americascup |
| Influencers | 9% | BTS, Jungkook, bag, eco leather, oversize, shirt, lamodechief, wkorea, vlive |

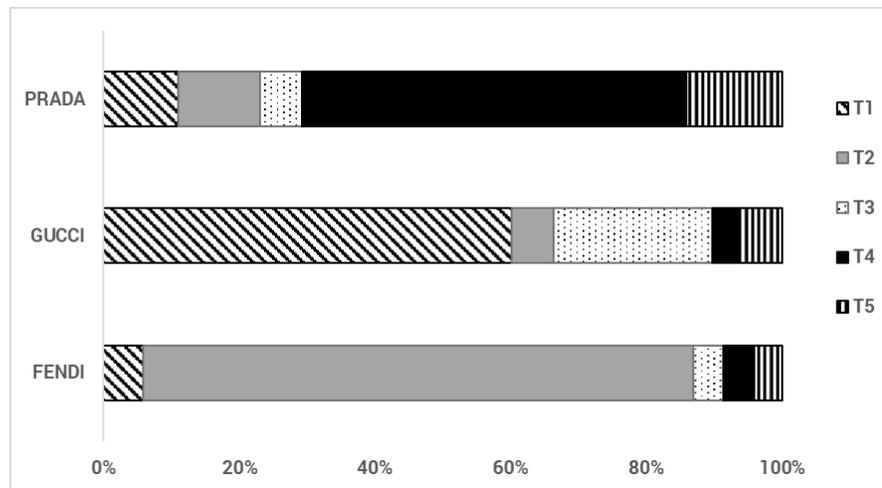

Figure 4. Brand-topic associations.

The first and most relevant topic is about the collaboration of Kai with Gucci, as Kai is consistently associated with this brand. The second topic is about well-known fashion brands, frequently mentioned together in tweets, which likely represents the messages of online shops or commercial offers. Topic 3 is about the movie "House of Gucci". The fourth topic is about social and sport events, while the last one is focused on the influence of the South Korean group BTS on fashion trends. Within this topic, there are words that have to do with media tools used by the band, such as the V LIVE video streaming application, or the fashion magazine W Korea. The band acts as an influencer and promotes fashion brands, by showing their outfits and through social media posts.

We then looked at how brands are associated with words that are related to several language dimensions, including the "affective" dimension, the "masculine" and "feminine" dimensions, and dimensions that identify a language focus on the present, future or past [8]. We provide an example in Figure 5, where we selected a few options among those available.

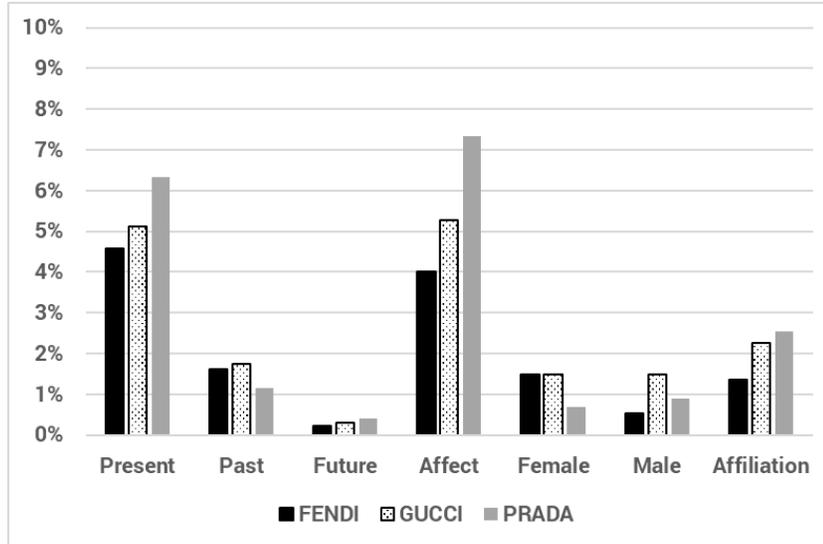

**Figure 5.** Example of brand image dimensions.

Not surprisingly, tweets are mostly related to events that happen in the present. The image of Gucci includes more masculine than feminine terms, whereas the opposite is true for Fendi. Certainly, this is affected by the week events (Kai ambassadorship of the brand) and might not be generalized to other timeframes. When looking at words that signal an affective process (e.g. "happy", "cry") we notice that Twitter users describe the brands in a rather emotional way [13]. This dimension is significantly higher for Prada, as well as the dimension of affiliation, typically communicated through messages of trust, liking, friendliness and companionship [14].

## 3 Conclusions

Gaining a deeper understanding of semantic brand importance and image can change business decisions and increase the ability to make predictions based on Big Data.

Brand managers can benefit from combining traditional surveys, which are often associated with low response rates, with the analysis of online forums, digital news, and social media posts. This can help them discover what people think about their offerings, and what promotional campaigns are generating a higher return on the investment.

Using the SBS BI app, and extending the period of observation to several weeks before, during and after marketing events, can help assess the brand equity at a more granular level. Instead of analyzing the "reported perception" of a brand from consumers filling out a survey, exploring how a brand is associated to others in the online discourse increases the objectivity of the analysis and provides a real time metric of performance, as it takes into account the 'spontaneous' discourse of stakeholders. By looking at social media posts, brand managers and campaign managers can understand consumers' opinions and emotions and adjust their initiatives right on time.